\documentclass[a4paper,aps,prd,10pt,preprintnumbers,showpacs,twocolumn,superscriptaddress,nofootinbib,amsmath,amssymb]{revtex4-1}
\usepackage[dvips]{graphics}
\usepackage[utf8]{inputenc}
\usepackage[T1]{fontenc}
\usepackage{cmap}
\def\imo{i}

\def\Order#1{{\cal O}\left(#1\right)}
\def\K{{\cal K}}

\begin{document}
\title{Analytic formula for quasinormal modes in the near-extreme Kerr-Newman-de Sitter spacetime governed by a non-Pöschl-Teller potential}
\author{M. S. Churilova}\email{wwrttye@gmail.com}
\affiliation{Research Centre for Theoretical Physics and Astrophysics, Institute of Physics, Silesian University in Opava, Bezručovo náměstí 13, CZ-74601 Opava, Czech Republic}
\author{R. A. Konoplya} \email{roman.konoplya@gmail.com}
\affiliation{Research Centre for Theoretical Physics and Astrophysics, Institute of Physics, Silesian University in Opava, Bezručovo náměstí 13, CZ-74601 Opava, Czech Republic}
\author{A. Zhidenko}\email{olexandr.zhydenko@ufabc.edu.br}
\affiliation{Research Centre for Theoretical Physics and Astrophysics, Institute of Physics, Silesian University in Opava, Bezručovo náměstí 13, CZ-74601 Opava, Czech Republic}
\affiliation{Centro de Matemática, Computação e Cognição (CMCC), Universidade Federal do ABC (UFABC),\\ Rua Abolição, CEP: 09210-180, Santo André, SP, Brazil}

\begin{abstract}
Quasinormal modes of scalar, electromagnetic, and gravitational fields in the extreme Schwarzschild-de Sitter background are known to be expressed in analytic form as eigenvalues of the Pöschl-Teller wavelike equation. We show that perturbations of fermionic fields (given by Dirac and Rarita-Schwinger equations) do not lead to the Pöschl-Teller effective potential. Nevertheless,  using the Frobenius method we find quasinormal modes analytically in this case as well. We write down the analytical formula for quasinormal frequencies of the near-extreme Schwarzschild-de Sitter black holes, which is valid for both bosonic and fermionic fields. We further extend the analysis to the case of charged rotating black holes and find a general analytical formula for quasinormal modes of the fields of various spin for the near extreme Kerr-Newman-de Sitter spacetime.
\end{abstract}
\maketitle

\section{Introduction}
Quasinormal modes are proper oscillation frequencies of gravitational or matter fields around black holes which depend only upon the black hole parameters and not on the way of perturbation of the system \cite{Kokkotas:1999bd,Konoplya:2011qq}.
One of the earliest and apparently the most elegant and simple calculation of quasinormal modes for Schwarzschild black holes was performed by Bahram Mashhoon~\cite{Mashhoon} via using the approximation of the exactly solvable eigenvalue problem given by the Pöschl-Teller wavelike equation. Later this approach was extended to the case of Reissner-Nordström and slowly rotating Kerr black holes \cite{Ferrari:1984zz}. The accuracy of the method, however, was insufficient, because it depends on how well the effective potential of the black hole fits the Pöschl-Teller potential.

Even though the study of quasinormal modes was concentrated around asymptotically flat solutions, the case of asymptotically de Sitter black holes has  also been considered in a great number of papers \cite{Zhidenko:2003wq,Suzuki:1998vy,Konoplya:2007zx,Yoshida:2010zzb,Suzuki:1999nn,Panotopoulos:2020mii,Churilova:2020mif,Tattersall:2018axd,Dey:2018cws,Panotopoulos:2018hua,Jansen:2017oag,Breton:2017hwe,Zhang:2014xha,Cuyubamba:2016cug,Fernando:2016ftj,Zhang:2014xha,Konoplya:2004uk,Konoplya:2007jv,Giammatteo:2005vu}, especially recently, in the context of Strong Cosmic Censorship \cite{Hod:2018lmi,Cardoso:2017soq,Dias:2018etb}. Mostly, calculations of quasinormal modes for four-dimensional black holes are done with the help of numerical or semianalytical methods, because the exact solution of the wavelike equation cannot be found as a rule. Nevertheless, there is one exception: the Schwarzschild-de Sitter black hole, when the cosmological horizon approaches the event horizon. In this regime of the near-extreme Schwarzschild-de Sitter spacetime the wavelike equation approaches the Pöschl-Teller one and, therefore, the analytical formula for quasinormal modes of scalar electromagnetic and gravitational fields can be deduced \cite{Cardoso:2003sw}. Later this formula for bosonic fields was extended to the case of higher-dimensional Reissner-Nordström-de Sitter black holes \cite{Molina:2003ff}. More general forms of the Pöschl-Teller-like potentials have been recently considered in \cite{Cardona:2017scd}.

Although from the point of view of current astrophysical observations we are able to detect only quasinormal modes of the gravitational field, there is considerable interest in perturbations of fundamental fields of other spin as well. A realistic black hole is not an isolated object, but it exists in the astrophysical environment and is accompanied by radiation phenomena of fields of all sorts. The decay of fermionic fields have also been extensively studied in the literature \cite{Jing:2003wq,Chen:2005rm,Chang:2005ki,Jing:2005dt,Jing:2005pk,Jing:2005cb,Li:2013fka,Varghese:2014xaa,Wahlang:2017zvk,Blazquez-Salcedo:2017bld,Gonzalez:2018xrq,Zinhailo:2019rwd,Wongjun:2019ydo,Aragon:2020teq,Chen:2021rty}, because the Dirac field can describe neutrinos, while the Rarita-Schwinger field corresponds to gravitinos \cite{Shu:2005qv,Zhang:2006bc}. After all, complete study of decay of all the fundamental fields allows one to understand the role of the spin field parameter in the processes of radiation.

In the present work we concentrate on the limit in which analytical expressions for quasinormal modes can be obtained. We will show here that perturbation equations for fermions do not reduce to the Pöschl-Teller wavelike equation, unlike the case of boson fields. Nevertheless, using the continued fractions we will obtain the analytical formula for quasinormal modes of the spin $1/2$ (Dirac) and $3/2$ (Rarita-Schwinger) fields in the regime of the near-extreme Schwarzschild-de Sitter black hole. We will also generalize this analytical treatment to the case of perturbations of scalar, electromagnetic, gravitational, Dirac and Rarita-Schwinger fields in the Kerr-Newman-de Sitter spacetime, allowing, unlike \cite{Ferrari:1984zz}, for the rotation and electric charge.

The paper is organized as follows. Section~\ref{sec:SdS} is devoted to a description of the Schwarzschild-de Sitter geometry in the near extreme limit. Section~\ref{sec:limit} deals with perturbation equations in the near extreme regime. In Section~\ref{sec:QNMs} we deduce an analytical formula for quasinormal modes of half-integer spin in the Schwarzschild-de Sitter spacetime, while in Section~\ref{sec:KNdS}, the analytical formula for quasinormal modes of  spin $0$, $1/2$, $1$, $3/2$, $2$ fields is deduced for the near extreme Kerr-Newman-de Sitter black hole. Finally, in Section~\ref{sec:conclusions} we summarize the obtained results and discuss a number of open questions.

\section{Schwarzschild-de Sitter geometry in the near extreme limit}\label{sec:SdS}
The line element of the Schwarzschild-de Sitter black hole can be written in the following form:
\begin{eqnarray}
  ds^2&=&-f(r)dt^2+\frac{dr^2}{f(r)}+r^2\left(d\theta^2+\sin^2\theta\,d\phi^2\right),\nonumber\\
  \label{metric}
\end{eqnarray}
where
\begin{equation}\label{metricfunc}
f(r)=1-\frac{2M}{r}-\frac{\Lambda r^2}{3}.
\end{equation}
Here $M>0$ is the black-hole mass and $\Lambda>0$ is the cosmological constant.

The function $f(r)$ has two positive roots, the event horizon $r_e$ and the cosmological horizon $r_c$, and can be expressed in the following form:
\begin{equation}\label{fhorizons}
  f(r)=\frac{\Lambda}{3r}(r_c-r)(r-r_e)(r+r_e+r_c).
\end{equation}

By comparing (\ref{fhorizons}) and (\ref{metricfunc}) one can express $M$ and $\Lambda$ in terms of $r_e$ and $r_c$ as follows:
\begin{equation}
 M =  \frac{r_er_c(r_e+r_c)}{2(r_e^2+r_er_c+r_c^2)}, \qquad  \Lambda = \frac{3}{r_e^2+r_er_c+r_c^2}.
\end{equation}

In this paper we consider the near-extreme Schwarzschild-de Sitter black hole, i.~e., the regime in which
\begin{equation}
r_c-r_e\ll r_e.
\end{equation}
In this case it is convenient to use the value of the surface gravity $\kappa_e$ as a small parameter,
\begin{eqnarray}\label{surfacegravity}
  \kappa_e\equiv \frac{f'(r_e)}{2}&=&\frac{\Lambda(r_c-r_e)(2r_e+r_c)}{6r_e}\\\nonumber
  &=&\frac{(r_c-r_e)}{2r_e^2}+\Order{r_c-r_e}^2.
\end{eqnarray}

We notice that
\begin{equation}\label{MLexpr}
M = \frac{r_e}{3}+\Order{\kappa_e}, \qquad \Lambda = \frac{1}{r_e^2}+\Order{\kappa_e},
\end{equation}
and the metric function (\ref{fhorizons}) takes the following form:
\begin{equation}\label{fexpr}
f(r)=\frac{(r_c-r)(r-r_e)}{r_e^2}+\Order{\kappa_e}^3.
\end{equation}

The tortoise coordinate takes the following simple form:
\begin{equation}\label{tortoise}
  r_*\equiv\int\frac{dr}{f(r)}=\frac{1}{2\kappa_e}\left(\ln\left(\frac{r-r_e}{r_c-r}\right)+\Order{\kappa_e}\right).
\end{equation}

Therefore, for the near-extreme Schwarzschild-de Sitter black hole one can find a closed form for $r$ in terms of $r_*$,
\begin{equation}\label{radialexpr}
  r=\frac{r_e+r_c \exp(2\kappa_er_*)}{1 + \exp(2\kappa_er_*)}+\Order{\kappa_e}.
\end{equation}

\section{Linear perturbations in the near-extreme limit}\label{sec:limit}
For static and spherically symmetric spacetime we can use the stationary ansatz and expansion into spherical harmonics for massless scalar ($s=0$), Dirac ($s=\pm1/2$), and electromagnetic ($s=1$) fields, as well as for linear gravitational perturbations ($s=2$) of odd end even parities,
$$\varphi,\Phi,A_{\mu},\delta g_{\mu\nu} \propto e^{-\imo\omega t}S(\theta,\phi)\Psi(r).$$
Then, the master equation for the radial part can be reduced to the wavelike form,
\begin{equation}\label{wavelike}
\left(\frac{d^2}{dr_*^2}+\omega^2-V_s(r)\right)\Psi=0,
\end{equation}
with the effective potentials (see, e.g., \cite{Regge:1957td,Brill:1957fx,Khanal:1983vb,Konoplya:2011qq} and references therein)
\begin{eqnarray}\label{s0}
&&V_0 = f(r)\left(\frac{\ell(\ell+1)}{r^2} + \frac{f'(r)}{r}\right),
\\\label{s1}
&&V_1 = f(r)\frac{\ell(\ell+1)}{r^2},
\\\label{s2}
&&V^{(o)}_2 = f(r)\left(\frac{\ell(\ell+1)}{r^2} - \frac{6M}{r^3}\right),
\\\nonumber
&&V^{(e)}_{2} = f(r)\Biggl\{\frac{\ell(\ell+1)}{r^2} - \frac{6M}{r^3}-\frac{24M^2\left(\Lambda r^2-3M/r\right)}{(6M + (\ell-1)(\ell+2)r)^2}
\\\nonumber&&\quad-\frac{24M(r-3M)}{(6M + (\ell-1)(\ell+2)r)^2}\left((\ell-1)(\ell+2)+\frac{3M}{r}\right)\Biggr\},
\\\label{sh}
&&V_{\pm1/2} = \frac{(\ell+1/2)^2 f(r)}{r^2}\pm (\ell+1/2)f(r)\frac{d}{dr}\frac{\sqrt{f(r)}}{r},
\end{eqnarray}
where the multipole number takes (half-)integer values, $\ell=|s|,|s|+1,|s|+2,\ldots$. While the deduction of the effective potentials for perturbations of the boson fields is well known in the literature (see for example \cite{Regge:1957td,Kokkotas:1999bd,Konoplya:2011qq} and references therein), the case of fermion fields is less known. Therefore, for completeness, we briefly review this deduction in the appendices.

Substituting (\ref{MLexpr}) and (\ref{fexpr}) into (\ref{s0}), (\ref{s1}), and (\ref{s2}), and taking into account (\ref{radialexpr}), we find that the potentials for $s=0,1,2$ approach the Pöschl-Teller one,
\begin{equation}
  V_s(r_*)=\kappa_e^2\frac{(\ell+s)(\ell+1-s)}{\cosh^2(\kappa_er_*)}+\Order{\kappa_e}^3,
\end{equation}
for which the quasinormal spectrum is known \cite{Ferrari:1984zz}. In this way it was shown that for the fields of the integer spin we have \cite{Cardoso:2003sw}
\begin{eqnarray}\label{integerspin}
  \frac{\omega}{\kappa_e}&=&\pm\sqrt{(\ell+s)(\ell+1-s)-\frac{1}{4}}
  \\\nonumber&&
  -\left(n+\frac{1}{2}\right)\imo +\Order{\kappa_e}, \qquad n=0,1,2,3\ldots.
\end{eqnarray}

However, for the half-integer spin, the effective potential is not the Pöschl-Teller like when $\kappa_e\to0$. In particular, by substituting (\ref{MLexpr}), (\ref{fexpr}), and (\ref{radialexpr}) into (\ref{sh}), one can find that the effective potential for the massless Dirac field takes the following form\footnote{We note that, when the cosmological horizon approaches the event horizon, one cannot consider $\sinh(\kappa_er_*)=\Order{\kappa_e}$ since $r_*$ is not a constant: Whatever small $\kappa_e$ is, the value of the hyperbolic sine goes from $-\infty$ to $\infty$ between the horizons. For the same reason one cannot take $\cosh(\kappa_er_*)=\Order{1}$.}:
\begin{equation}\label{Dirac}
  V_{\pm1/2}(r_*)=\kappa_e^2\left(\ell+\frac{1}{2}\right)\frac{\ell+\frac{1}{2}\mp\sinh(\kappa_er_*)}{\cosh^2(\kappa_er_*)}+\Order{\kappa_e}^3.
\end{equation}
Similarly, the effective potential for the Rarita–Schwinger field is reduced to the non-Pöschl-Teller one in the near-extreme limit of the Schwarzschild-de Sitter black hole (see Appendix~\ref{sec:halfintegerspin}).

\section{Quasinormal spectrum of the half-integer spin fields}\label{sec:QNMs}
Using the Newman-Penrose formalism it is possible to separate perturbation equations for massless fields of any spin in the background of the Kerr-Newman-de Sitter geometry \cite{Newman:1961qr}. After some algebra, the perturbation equations can be reduced to the master equation for the radial part of the conformally coupled scalar ($s=0$), massless Dirac ($s=\pm1/2$), Maxwell ($s=\pm1$), Rarita–Schwinger field ($s=\pm3/2$), and linear gravitational perturbations ($s=\pm2$) of the Schwarzschild-de Sitter black holes can be written in the following form \cite{Suzuki:1998vy} (see Appendix~\ref{sec:master}):
\begin{eqnarray}\label{mastereq}
\Biggl\{\left(\frac{f(r)}{r^2}\right)^{1-s}\frac{d}{dr}\left(\frac{f(r)}{r^2}\right)^{s+1}\frac{d}{dr} + \omega^2 \qquad\qquad\qquad
\\\nonumber
+\imo s \omega\frac{2f(r)-rf'(r)}{r} -\frac{2(s+1)(2s+1)\Lambda f(r)}{3}
\\\nonumber
-\frac{f(r)(\ell-s)(\ell+1+s)}{r^2}\Biggr\} R(r) &=& 0.
\end{eqnarray}
Here $\ell=|s|,|s|+1,|s|+2,\ldots$ is the (half-)integer multipole number and $\omega$ is the frequency.

Following \cite{Konoplya:2007zx}, we represent solutions to (\ref{mastereq}) using the Frobenius expansion,
\begin{equation}\label{Frobenius}
  R(r)=\left(1-\frac{r_e}{r}\right)^{-s-\imo\omega/\kappa_e}\frac{e^{\imo\omega r_*}}{r^{2s+1}}\sum_{k=0}^{\infty}b_k\left(\frac{1-r_e/r}{1-r_e/r_c}\right)^k.
\end{equation}
The coefficients $b_k$ satisfy the three-term recurrence relation,
\begin{eqnarray}\label{recurrence}
  && c_{0,1}(\omega)b_1+c_{1,1}(\omega)b_0=0,
  \\\label{recurrence-three}
  && c_{0,n}(\omega)b_n+c_{1,n}(\omega)b_{n-1}+c_{2,n}(\omega)b_{n-2}=0,
\end{eqnarray}
with
\begin{eqnarray}\nonumber
c_{0,n}(\omega)&=&n r_c(r_c-r_e)(r_c+2r_e)^2\biggl\{2\imo\omega r_e(r_e^2+r_er_c+r_c^2)
\\\nonumber&&-(r_c-r_e)(r_c+2r_e)(n-s)\biggr\}=\Order{\kappa_e}^2,
\\\nonumber
c_{1,n}(\omega)&=& \ell(\ell+1)(r_c-r_e)(r_c+2r_e)^2(r_c^3-r_e^3)
\\\nonumber&&+(1-s^2)r_c(r_c-r_e)(r_c+2r_e)^2(r_c^2-r_e^2)
\\\nonumber&&-(2r_c^2+2r_cr_e-r_e^2)\Biggl\{4\omega^2r_e^2(r_e^2+r_er_c+r_c^2)^2
\\\nonumber&&+2\imo\omega r_e(2n-1)(r_c^3-r_e^3)(r_c+2r_e)
\\\nonumber&&-n(n-1)(r_c-r_e)^2(r_c+2r_e)^2\Biggr\}
\\\nonumber
&=&108r_e^8\Biggl\{\kappa_e^2 \left(\ell(\ell+1)+n(n-1)+\frac{2}{3}(1-s^2)\right)
\\\nonumber&&-2\imo\omega\kappa_e (2n-1)-\omega^2\Biggr\}+\Order{\kappa_e}^3,
\\\nonumber
c_{2,n}(\omega)&=&-r_e^2(r_c^2-r_e^2)(r_e^2+r_er_c+r_c^2)^2\times
\\\nonumber&\times&\left(2\imo\omega-\frac{(n +s-1)(r_c-r_e)(r_c+2r_e)}{r_e(r_e^2+r_er_c+r_c^2)}\right)\times
\\\nonumber&\times&\left(2\imo\omega-\frac{(n-1)(r_c-r_e)(r_c+2r_e)}{r_e(r_e^2+r_er_c+r_c^2)}\right)=\Order{\kappa_e}^3,
\end{eqnarray}
where we took into account that $\omega=\Order{\kappa_e}$.

One can express $b_1$ using either (\ref{recurrence}) or (\ref{recurrence-three}) via the following relation:
\begin{equation}\label{coeffcond}
\frac{b_1}{b_0}=-\frac{c_{1,1}(\omega)}{c_{0,1}(\omega)} = -\frac{c_{2,2}(\omega)}{c_{1,2}(\omega)-}\,\frac{c_{0,2}(\omega)c_{2,3}(\omega)}{c_{1,3}(\omega)-}\ldots\,.
\end{equation}
Equation~(\ref{coeffcond}) with the infinite continued fraction is satisfied if and only if the Frobenius series (\ref{Frobenius}) converges. By inverting the continued fraction $n$ times, we obtain the following equivalent equation \cite{Leaver:1985ax,Nollert:1993zz}:
\begin{eqnarray}\label{contfrac}
c_{1,n+1}(\omega)-\frac{c_{2,n}(\omega)c_{0,n-1}(\omega)}{c_{1,n-1}(\omega)-}\ldots\,\frac{c_{2,2}(\omega)c_{0,1}(\omega)}{c_{1,1}(\omega)}=\qquad
\\\nonumber=\frac{c_{0,n+1}(\omega)c_{2,n+2}(\omega)}{c_{1,n+2}(\omega)-}\frac{c_{0,n+2}(\omega)c_{2,n+3}(\omega)}{c_{1,n+3}(\omega)-}\ldots\,,
\end{eqnarray}
which must be solved with respect to the quasinormal frequency $\omega$.

For the near-extreme Schwarzschild-de Sitter black hole equation (\ref{contfrac}) takes the simpler form,
\begin{equation}\label{simplified}
c_{1,n+1}(\omega)=\Order{\kappa_e}^3,
\end{equation}
and can be solved analytically,
\begin{eqnarray}\label{allspin}
\frac{\omega}{\kappa_e}&=&\pm\sqrt{(\ell+s)(\ell+1-s)+\frac{(2s-1)(2s-5)}{12}}
\\\nonumber&&-\left(n+\frac{1}{2}\right)\imo+\Order{\kappa_e}, \qquad n=0,1,2,3,\ldots.
\end{eqnarray}
Notice that formula (\ref{allspin}) coincides with (\ref{integerspin}) for $s=1$ and $s=2$. For $s=0$ the spectrum (\ref{allspin}) corresponds to the quasinormal modes of the conformally coupled scalar field, while (\ref{integerspin}) was derived for the minimally coupled scalar field.

The simplification of Eq.~(\ref{contfrac}) occurs because the coefficients $c_{2,n}(\omega)$ in the near-extreme limit are negligible compared to $c_{0,n}(\omega)$ and $c_{1,n}(\omega)$. As a result, Eq.~(\ref{recurrence-three}) becomes the two-terms recurrence relation and Eq.~(\ref{simplified}) implies that the infinite series in~(\ref{Frobenius}) is reduced to a finite sum, when $\omega$ is a quasinormal frequency. The Frobenius series for the Kerr-de Sitter black-hole perturbations, proposed in \cite{Yoshida:2010zzb}, leads to the same simplification. Therefore, one can easily obtain a generalization of (\ref{allspin}) for rotating black holes (see Sec.~\ref{sec:KNdS}). It is interesting to note that the expansion in terms of the hypergeometric functions \cite{Suzuki:1999nn} is not simplified in this case, leading to a three-terms relation when the event horizon approaches the de Sitter one.

\begin{table}
\begin{tabular}{p{10em}c}
\hline
$(r_c-r_e)/r_e$ & $\omega/\kappa_e$    \\
\hline
$0.000900$ & $1.999401 - 0.499853\imo$ \\
$0.000265$ & $1.999823 - 0.499958\imo$ \\
$0.000171$ & $1.999886 - 0.499974\imo$ \\
$0.000152$ & $1.999899 - 0.499977\imo$ \\
$0.000142$ & $1.999906 - 0.499979\imo$ \\
\hline
\end{tabular}
\caption{The fundamental quasinormal mode of the Dirac field, $s=1/2$, $\ell=3/2$, obtained by the WKB method with Padé approximants, $P_{5/1}$. The corresponding value, given by our analytical formula, is $2-0.5\imo$.}\label{tabl:converge}
\end{table}

Our analytical formula can be checked with the higher-order WKB (Wentzel–Kramers–Brillouin) formula of Will and Schutz~\cite{Schutz:1985km}, which was extended to higher orders in \cite{Iyer:1986np,Konoplya:2003ii,Matyjasek:2017psv} and made even more accurate by the usage of the Padé approximants in \cite{Matyjasek:2017psv,Hatsuda:2019eoj}.
The higher-order WKB formula reads \cite{Konoplya:2019hlu}
\begin{eqnarray}
 \omega^2&=&V_0+A_2(\K^2)+A_4(\K^2)+A_6(\K^2)+\ldots \\\nonumber
&-& \imo \K\sqrt{-2V_2}\left(1+A_3(\K^2)+A_5(\K^2)+A_7(\K^2)+\ldots\right),
\end{eqnarray}
where $\K$ takes half-integer values, $\K=n+\frac{1}{2}$. The corrections $A_k(\K^2)$ of order $k$ to the eikonal formula are polynomials of $\K^2$ with rational coefficients and depend on the values of higher derivatives of the potential $V(r)$ in its maximum. In order to increase accuracy of the WKB formula, we follow Matyjasek and Opala \cite{Matyjasek:2017psv} and use Padé approximants. The Padé approximants $P_{\tilde{n}/\tilde{m}}$ are defined in Eq.~(21)~of~\cite{Konoplya:2019hlu}. In Table~\ref{tabl:converge} we show that the fundamental ($n=0$) quasinormal mode of the Dirac field ($s=1/2$) for $\ell=3/2$, obtained by the WKB method with Padé approximants, approaches the limit
$$\omega/\kappa_e \to 2-0.5\imo,$$
which coincides with the analytical result (\ref{allspin}).

\bigskip\bigskip\bigskip\bigskip

As a by-product, let us consider a more general equation
\begin{equation}\label{modifiedPT}
  \left(\frac{d^2}{dx^2}+\omega^2-\frac{A^2+A\alpha\sinh(\alpha x)}{\cosh^2(\alpha x)}\right)\Psi(x)=0,
\end{equation}
  where $\alpha$ and $A$ are real nonzero parameters.
Since (\ref{allspin}) for $s=1/2$ gives us the quasinormal spectrum of the wavelike equation (\ref{wavelike}) with the effective potential (\ref{Dirac}), we conclude that
\begin{equation}\label{modifiedPTsol}
  \omega=\pm A-|\alpha|\left(n+\frac{1}{2}\right)\imo, \qquad n=0,1,2,3,\ldots
\end{equation}
are quasinormal modes for the general wave equation (\ref{modifiedPT}).

It is clear that the effective potential in (\ref{modifiedPT}) always has a negative gap for sufficiently large values of $x$, either positive (for $A<0$) or negative (for $A>0$). Nevertheless, as can be seen from the above analytic formulas for quasinormal modes, there are no unstable (growing) modes in the spectrum.

\section{Quasinormal modes of the near-extreme Kerr(-Newman)-de Sitter black holes}\label{sec:KNdS}
Here we extend our analysis to the case of the Kerr(-Newman)-de Sitter spacetime, describing a charged and rotating black hole. The line element for this case has the form:
\begin{eqnarray}\label{KNdSmetric}
ds^2 &=& -\frac{\Delta_r(dt-a\sin^2\theta d\phi)^2}{(1+\Lambda a^2/3)^2(r^2+a^2\cos^2\theta)}
\\\nonumber
&&
\qquad+\frac{\Delta_\theta \sin^2\theta\left(adt-(r^2+a^2)d\phi\right)^2}{(1+\Lambda a^2/3)^2(r^2+a^2\cos^2\theta)}
\\\nonumber
&&
\qquad\qquad+(r^2+a^2\cos^2\theta)\left(\frac{dr^2}{\Delta_r}+\frac{d\theta^2}{\Delta_\theta}\right),
\end{eqnarray}
where
\begin{eqnarray}
\Delta_r&=&(r^2+a^2)\left(1-\frac{\Lambda r^2}{3}\right)-2Mr+Q^2\\\nonumber
&=&\frac{\Lambda}{3}(r_c-r)(r-r_e)(r-r_i)(r+r_c+r_e+r_i), \\
\Delta_\theta&=&1+\frac{\Lambda a^2}{3}\cos^2\theta.
\end{eqnarray}
Here $a$ is the rotation parameter and $Q$ is the electric charge. The latter can be expressed in terms of the inner horizon $r_i$, for which $0<r_i<r_e<r_c$.

For the charged black hole the background electromagnetic field is given by the components of the one-form
\begin{equation}
A_{\mu}dx^{\mu}=-\frac{Qr}{r^2+a^2\cos^2\theta}\cdot\frac{dt-a\sin^2\theta d\phi}{(1+\Lambda a^2/3)^2}.
\end{equation}

The surface gravity is given by the following relation:
\begin{eqnarray}\nonumber
  \kappa_e&\equiv&\frac{(r_c - r_e) (r_e - r_i) (r_c + 2 r_e + r_i)}{2 (a^2 + r_e^2) (2 a^2 + r_c^2 + r_e^2  + r_i^2 + r_e r_i + r_c r_e + r_c r_i)}
  \\
  &=&\Order{r_c-r_e}.
\end{eqnarray}

For the uncharged black hole ($Q=0$) the inner horizon can be expressed in terms of the rotation parameter,
\begin{eqnarray}\nonumber
  r_i&=&\frac{r_e+r_c}{2}\left(\sqrt{1+\frac{4a^2}{r_cr_e-a^2}-\frac{4a^2}{(r_e+r_c)^2}}-1\right)
  \\\label{innerhor}
     &=&\frac{r_e^2+a^2}{\sqrt{r_e^2-a^2}}-r_e+\Order{\kappa_e}.
\end{eqnarray}

Omitting the deduction of the perturbation equation, which can be found in Appendix~\ref{sec:master}, we proceed to the analysis of the wavelike equation.
The corresponding recurrence-relation coefficients have been derived in \cite{Yoshida:2010zzb}. Taking the limit $\kappa_e\to 0$, we obtain
\begin{equation}\label{KNdSeq}
\beta_r=\Order{\kappa_e},
\end{equation}
where $\beta_r$ is given by (54) in \cite{Yoshida:2010zzb}. Equation~(\ref{KNdSeq}) can be solved analytically as follows:
\begin{eqnarray}\nonumber
  \omega&=&m\Omega_e\pm\kappa_e\sqrt{C_1\lambda+C_2^2m^2\Omega_e^2r_e^2+\frac{(2s-1)(2s-C_3)}{C_4}}
  \\\label{KNdSspectrum}
  &&-C_2m\Omega_e r_e\kappa_e -\imo\left(n+\frac{1}{2}\right)\kappa_e+\Order{\kappa_e}^2,
\end{eqnarray}
where $\Omega_e\equiv\dfrac{a}{a^2+r_e^2}$ is the angular velocity at the event horizon, $m=-\ell,-\ell+1,\ldots,\ell-1,\ell$ is the azimuthal (half-)integer number, $n=0,1,2,3,\ldots$ is the overtone number, and $C_1$, $C_2$, $C_3$, and $C_4$ are positive constants,
\begin{equation}
\begin{array}{rcl}
  C_1&=&\dfrac{a^2 + 3 r_e^2 + 2 r_e r_i + r_i^2}{(r_e - r_i) (3 r_e + r_i)},
  \\
  C_2&=&2\dfrac{2 a^2 + 3 r_e^2 + 2 r_e r_i + r_i^2}{(r_e - r_i) (3 r_e + r_i)},
  \\
  C_3&=&\dfrac{5 r_e^2 + 2 r_e r_i + r_i^2}{(r_e + r_i)^2},
  \\
  C_4&=&4 \dfrac{(r_e - r_i) (3 r_e + r_i)}{(r_e + r_i)^2}.
\end{array}
\end{equation}
When the charge $Q$ and rotation parameter $a$ go to zero, the general formula (\ref{KNdSspectrum}) is reduced to (\ref{allspin}).
Notice that the expression for $v_0$ in \cite{Yoshida:2010zzb} has misprints and uses such dimensional units that $\Lambda=3$. The accurate three-terms recurrence relation as well as derivation of the formula (\ref{KNdSspectrum}) are available in the Wolfram Mathematica\textregistered{} ancillary file \cite{acillary}.

The separation constant $\lambda$ can be calculated for any $\omega$ (see Appendix~\ref{sec:master}). In the near-extreme limit of the slowly rotating Kerr-de Sitter black hole ($Q=0$), substituting (\ref{KNdSspectrum}) into (\ref{angularslow}), we find
\begin{eqnarray}
\lambda&=&(\ell+s)(\ell+1-s)\left(1-\frac{a^2}{3r_e^2}\right)
\\\nonumber&&+\frac{2a^2}{r_e^2}\Biggl(\frac{((\ell+1)^2 - m^2) ((\ell+1)^2 - s^2)^2}{(2\ell+1)(\ell+1)(2\ell+3)}-\frac{2m^2}{3}
\\\nonumber&&-\frac{2m^2s^2}{3\ell(\ell+1)}-\frac{(\ell^2 - m^2) (\ell^2 - s^2)^2}{(2\ell-1)\ell(2\ell+1)} \Biggr)+\Order{a^4,\kappa_e}.
\end{eqnarray}

Notice that the general expression (\ref{KNdSspectrum}) is valid for highly rotating black holes as well, although the separation constant $\lambda=\lambda(s,\ell,m,a,r_i,r_e=r_c)$ has not been found in analytic form in the that case.

\section{Conclusions}\label{sec:conclusions}
Quasinormal modes of the four-dimensional asymptotically flat or de Sitter black holes are found numerically as a rule. The exception is bosonic perturbations of the near extreme Schwarzschild-de Sitter black hole for which the wavelike equation takes the Pöschl-Teller form. In this case the quasinormal spectrum of the Schwarzschild-de Sitter black hole can be found analytically. Here we have shown that perturbations of fermionic (Dirac and Rarita-Schwinger) fields are not reduced to the Pöschl-Teller-like form. Nevertheless, using the Frobenius series, we have found the analytical formula for quasinormal modes of half-integer spin fields and, consequently, the general formula for spin $0$, $1/2$, $1$, $3/2$, $2$ fields. We have also generalized this analytical formula to the case of Kerr-Newman-de Sitter black hole. Our analysis can be extended to the case of a higher dimensional gravity, as well as to black holes in four and higher-dimensional modified gravities.

\begin{acknowledgments}
We acknowledge support of the grant 19-03950S of Czech Science Foundation (GAČR).
\end{acknowledgments}

\newpage

\appendix

\section{Effective potentials for the fermionic fields in the Schwarzschild-de Sitter background}\label{sec:halfintegerspin}
\subsection{Dirac field}
\def\dirac{{\slash \negthinspace \negthinspace \negthinspace \nabla}}
The massless Dirac equation
\begin{equation} \label{eq: Dirac equation}
 \imo \dirac \psi = 0, \qquad \psi\propto e^{-\imo\omega t},
\end{equation}
in the Schwarzschild-de Sitter background simplifies to the coupled system of partial differential equations \cite{Gibbons:1993hg,Das:1996we,Lopez-Ortega:2004vrh}
\begin{equation} \label{eq: Dirac equation simplified}
\begin{array}{rcl}
 -\imo\omega \psi_1 + \dfrac{d\psi_1}{dr_*} & =  W \psi_2,  \\
 -\imo\omega \psi_2 - \dfrac{d\psi_2}{dr_*} & = - W \psi_1,
\end{array}
\end{equation}
where $\psi_1$ and $\psi_2$ are the components of a two-dimensional spinor,
\begin{equation} \label{eq: W function general}
 W = \left( \ell+1/2 \right)\frac{\sqrt{f(r)}}{r},
\end{equation}
where $\ell$ is a positive half-integer number, $\ell=\frac{1}{2},\frac{3}{2},\frac{5}{2},\dots$.

One can rewrite Eqs.~(\ref{eq: Dirac equation simplified}) as follows:
\begin{equation} \label{eq: coupled equations Zplus Zminus}
\begin{array}{rcl}
\dfrac{dZ_+}{dr_*} + \imo\omega Z_- &= W Z_+ , \\
\dfrac{dZ_-}{dr_*} + \imo\omega Z_+ &= - W Z_- ,
\end{array}
\end{equation}
where
\begin{equation} \label{eq: definitions Z and tortoise}
Z_\pm = \psi_2 \pm \psi_1.
\end{equation}

From Eqs.~(\ref{eq: coupled equations Zplus Zminus}) one can obtain the wavelike equations for the functions $Z_\pm$
\begin{equation}
 \frac{d^2Z_\pm}{dr_*^2} + \omega^2 Z_\pm = V_{\pm1/2} Z_\pm ,
\end{equation}
where the effective potentials $V_{\pm1/2}$ are given by,
\begin{equation} \label{eq: potentials plus minus}
 V_{\pm1/2} = W^2 \pm \frac{dW}{dr_*}.
\end{equation}
The latter relation leads immediately to Eq.~(\ref{sh}).

\subsection{Rarita-Schwinger field}
It was shown in \cite{Chen:2019rob} that equations for the components of the Rarita-Schwinger field in the Schwarzschild-de Sitter background can be reduced to the form (\ref{eq: Dirac equation simplified}) as well. However, in this case both the function $W$ and the tortoise coordinate $r_*$ depend on $\omega$,
\begin{eqnarray}\label{modified-tortoise}
   dr_* &=& \frac{dr}{f(r)}\left[1 + \frac{f}{2\omega}\left(\frac{d}{d r}\frac{\mathcal{D}}{i\mathcal{B}} \right)\left(\frac{\mathcal{B}^{2}}{\mathcal{B}^{2}-\mathcal{D}^{2}} \right) \right]\; ,\\\nonumber
    W &=& \sqrt{\mathcal{D}^{2}-\mathcal{B}^{2}}\left[1 + \frac{f}{2\omega}\left(\frac{d}{d r}\frac{\mathcal{D}}{i\mathcal{B}} \right)\left(\frac{\mathcal{B}^{2}}{\mathcal{B}^{2}-\mathcal{D}^{2}} \right) \right]^{-1}\; ,
\end{eqnarray}
with
\begin{eqnarray}\nonumber
\mathcal{B} &=& \frac{i(\ell+1/2)\sqrt{f(r)}}{r}\left(1 + \frac{2M/r}{f(r) + \frac{1}{3}\Lambda r^2 - (\ell+1/2)^{2}}\right),\\
\mathcal{D}&=&-i\sqrt{\frac{\Lambda f(r)}{3}}\frac{2M/r}{f(r) + \frac{1}{3}\Lambda r^2 - (\ell+1/2)^{2}}\; ,
\end{eqnarray}
where the multipole number $\ell=\frac{3}{2},\frac{5}{2},\frac{7}{2},\dots$.

Thus, the effective potential for the Rarita-Schwinger field does not depend on $\omega$ only in the Schwarzschild limit \cite{Chen:2015jga}.

It is interesting to note that in the limit of the near-extreme Schwarzschild-de Sitter black hole Eq.~(\ref{modified-tortoise}) takes the form,
\begin{eqnarray}
dr_* &=& \frac{dr}{f(r)}+\Order{\kappa_e},\\\nonumber
W &=& \frac{\sqrt{f(r)}}{r}\cdot\sqrt{\ell^2+\ell-\frac{13}{12}}+\Order{\kappa_e}^2,
\end{eqnarray}
and the effective potential approaches
\begin{eqnarray}
  V_{\pm3/2}(r_*)&=&\kappa_e^2\sqrt{\ell^2+\ell-\frac{13}{12}}\frac{\sqrt{\ell^2+\ell-\frac{13}{12}}\pm\sinh(\kappa_er_*)}{\cosh^2(\kappa_er_*)}
  \nonumber\\&&+\Order{\kappa_e}^3.
\label{RaritaSchwinger}
\end{eqnarray}

The potential (\ref{RaritaSchwinger}) has the same form as the effective potential for the Dirac field (\ref{modifiedPT}). Therefore, from Eq.~(\ref{modifiedPTsol}) one can obtain the correct spectrum given by (\ref{allspin}) for $s=\pm\frac{3}{2}$.

\section{Master equations for the Kerr-Newman-de Sitter black-hole perturbations}\label{sec:master}
We use the Newman-Penrose formalism \cite{Newman:1961qr} in order to separate the perturbation equations in the background of the Kerr-Newman-de Sitter black hole (\ref{KNdSmetric}).

Following \cite{Suzuki:1998vy}, we introduce the following vectors as the null tetrad:
\begin{eqnarray}
l^\mu &=& \frac{1+\alpha}{\Delta_r}\left(r^2+a^2, \frac{\Delta_r}{1+\alpha}, 0,
a\right), \quad \alpha\equiv\frac{\Lambda a^2}{3},  \\
n^\mu &=& \frac{1+\alpha}{2(r^2+a^2\cos^2\theta)}\left(r^2+a^2, -\frac{\Delta_r}{1+\alpha}, 0, a\right), \nonumber \\
m^\mu &=& \frac{1+\alpha}{(r+\imo a\cos\theta)\sqrt{2\Delta_\theta}}
\left(\imo a\sin\theta, 0, \frac{\Delta_\theta}{1+\alpha},
\frac{\imo}{\sin\theta}\right),\nonumber \\
\bar{m}^\mu&=& \frac{1+\alpha}{(r-\imo a\cos\theta)\sqrt{2\Delta_\theta}}
\left(-\imo a\sin\theta, 0, \frac{\Delta_\theta}{1+\alpha},
-\frac{\imo}{\sin\theta}\right), \nonumber
\end{eqnarray}
and assume the following ansatz for all the scalar quantities:
\begin{equation}\label{ansatz}
  \Phi_s(t,r,\theta,\phi) \propto e^{-\imo\omega t+\imo m \phi}S_s(\theta)R_s(r).
\end{equation}

It is convenient to use the tetrad components of the derivative and the electromagnetic field as follows:
\begin{equation}
\begin{array}{rcl}
l^\mu \partial_\mu&=&{\cal D}_0, \\
n^\mu \partial_\mu&=&\displaystyle{-\frac{\Delta_r}{2(r^2+a^2\cos^2\theta)}{\cal D}_0^\dag}, \\
m^\mu \partial_\mu&=&\displaystyle{\frac{\sqrt{\Delta_\theta}}{\sqrt{2}(r+\imo a\cos\theta)}{\cal L}_0^\dag}, \\
\bar{m}^\mu \partial_\mu&=&\displaystyle{\frac{\sqrt{\Delta_\theta}}{\sqrt{2}(r-\imo a\cos\theta)}{\cal L}_0}, \\
l^\mu A_\mu&=&\displaystyle{-\frac{Qr}{\Delta_r}}, \\
n^\mu A_\mu&=&\displaystyle{-\frac{Qr}{2(r^2+a^2\cos^2\theta)}}, \\
m^\mu A_\mu&=&\bar{m}^\mu A_\mu=0,
\end{array}
\end{equation}
where we introduced
\begin{eqnarray}
{\cal D}_n&=&\partial_r-\frac{\imo(1+\alpha)K}{\Delta_r}
+n\frac{\partial_r \Delta_r}{\Delta_r}, \nonumber \\
{\cal D}_n^\dag&=&\partial_r+\frac{\imo(1+\alpha)K}{\Delta_r}
+n\frac{\partial_r \Delta_r}{\Delta_r},  \\
K&=&\omega(r^2+a^2)-am; \nonumber
\end{eqnarray}
\begin{eqnarray}
{\cal L}_n&=&\partial_\theta+\frac{(1+\alpha)H}{\Delta_\theta}
+n\frac{\partial_\theta(\sqrt{\Delta_\theta} \sin\theta)}
{\sqrt{\Delta_\theta} \sin\theta}, \nonumber \\
{\cal L}_n^\dag&=&\partial_\theta-\frac{(1+\alpha)H}{\Delta_\theta}
+n\frac{\partial_\theta(\sqrt{\Delta_\theta} \sin\theta)}
{\sqrt{\Delta_\theta} \sin\theta}, \\
H&=&-a\omega\sin\theta+\frac{m}{\sin\theta}.\nonumber
\end{eqnarray}

The gravitational-led perturbations of the Kerr-Newman-de Sitter black-hole can be expressed in terms of the Weyl tensor $C_{\mu\nu\lambda\sigma}$,
\begin{equation}\label{gravscalars}
\begin{array}{rcl}
\Phi_{+2}(t,r,\theta,\phi)&\equiv&C_{\mu\nu\lambda\sigma}l^{\mu}m^{\nu}l^{\lambda}m^{\sigma},\\
\Phi_{-2}(t,r,\theta,\phi)&\equiv&\dfrac{C_{\mu\nu\lambda\sigma}n^{\mu}\bar{m}^{\nu}n^{\lambda}\bar{m}^{\sigma}}{(r^2+a^2\cos^2\theta)^2},
\end{array}
\end{equation}
and the quantities
\begin{equation}\label{Maxwellscalars}
\begin{array}{rcl}
\Phi_{+1}(t,r,\theta,\phi)&\equiv&F_{\mu\nu}l^{\mu}m^{\nu},\\
\Phi_{-1}(t,r,\theta,\phi)&\equiv&\dfrac{F_{\mu\nu}\bar{m}^{\mu}n^{\nu}}{r^2+a^2\cos^2\theta},
\end{array}
\end{equation}
correspond to the perturbations due to the electromagnetic field, $F_{\mu\nu}\equiv\partial_{\mu}A_{\nu}-\partial_{\nu}A_{\mu}$.

The conformally coupled test scalar field obeys
\begin{equation}\label{conformalscalar}
  \frac{1}{\sqrt{-g}}\partial_{\mu}\sqrt{-g}g^{\mu\nu}\partial_{\nu}\Phi_0=\frac{R}{6}\Phi_0,
\end{equation}
where $R=4\Lambda$ is the scalar curvature.

In \cite{Suzuki:1998vy} it was shown that the perturbation equations for:
\begin{itemize}
\item the test scalar field $\Phi_0$ obeying the Klein-Gordon equation (\ref{conformalscalar});
\item the independent components of the massless Dirac field, $\Phi_{\pm1/2}$;
\item Maxwell field wave function $\Phi_{\pm1}$ defined in (\ref{Maxwellscalars});
\item the independent components of the Rarita-Schwinger field, $\Phi_{\pm3/2}$;
\item gravitational perturbations $\Phi_{\pm2}$ defined in (\ref{gravscalars});
\end{itemize}
allow for separation of the angular variables and, in the end, satisfy the Teukolsky equations \cite{Teukolsky:1974yv}
\begin{eqnarray} \label{eqn:Ss}
&& \left[ \sqrt{\Delta_\theta}{\cal L}_{1-s}^{\dag}\sqrt{\Delta_\theta} {\cal L}_s - 2(1+\alpha)(2s-1)a\omega\cos\theta \right. \\
&& \qquad\qquad \left. - 2\alpha(s-1)(2s-1)\cos^2\theta+\lambda \right]S_s(\theta) = 0, \nonumber \\
\label{eqn:Rs}
&& \left[ \Delta_r {\cal D}_1 {\cal D}_s^\dag +2(1+\alpha)(2s-1)\imo\omega r \right. \\ \nonumber
&& \qquad\qquad \left. -\frac{2\Lambda}{3}(s+1)(2s+1)r^2 - \lambda \right]R_s(r) = 0,
\end{eqnarray}
where $\lambda$ is the separation constant. Here $s=0,\pm1/2,\pm1,\pm3/2,\pm2$, which corresponds to the scalar, Dirac, Maxwell, Rarita-Schwinger, and gravitational perturbations, respectively.

After introducing the new variable, $x\equiv\cos\theta$, the coefficients of the angular Teukolsky equation (\ref{eqn:Ss}) become rational functions of $x$ with five regular singularities, one of which, corresponding to $x=\infty$, can be factored out. In this way, equation (\ref{eqn:Ss}) is reduced to the Heun’s equation, and the separation constant $\lambda$ can be computed for any given value of $\omega$. In particular, for the slowly rotating black hole, one can obtain Eq.~(4.18) of \cite{Suzuki:1998vy}, which has the following cumbersome form:
\begin{widetext}
\begin{eqnarray}\label{angularslow}
\lambda &=& (\ell + 1 - s)(\ell + s) +\alpha\left[ \ -(\ell + 1 - s)(\ell + s) + 2m^2 + \frac{2m^2s^2}{\ell(\ell+1)}-\ell^2 H(\ell)+(\ell+1)^2H(\ell+1) \ \right]
\\ &&
+\left\{-2m\left(1+\frac{s^2}{\ell(\ell+1)}\right) - 2m\alpha\left[1+\frac{s^2}{\ell(\ell+1)} -\left(1+\frac{s^2}{(\ell-1)(\ell+1)}\right)H(\ell) +\left(1+\frac{s^2}{\ell(\ell+2)}\right)H(\ell+1)\right] \right\} a\omega
\nonumber \\ && +\Bigg\{ H(\ell+1)-H(\ell) +\alpha\Bigg[H(\ell+1)-H(\ell)+2\Big((\ell+1)^2H(\ell+1)-\ell^2H(\ell)\Big)
\nonumber \\ &&
- \ell H^2(\ell)+(\ell+1)H^2(\ell+1)-\frac{H(\ell)H(\ell+1)}{\ell(\ell+1)} + 6m^2s^2\Big(\frac{H(\ell+1)}{\ell(\ell+1)^2(\ell+2)} -\frac{H(\ell)}{(\ell-1)\ell^2(\ell+1)}\Big)
\nonumber \\ &&
+4m^2s^4\Big(\frac{H(\ell+1)}{\ell^2(\ell+1)^2(\ell+2)^2} -\frac{H(\ell)}{(\ell-1)^2\ell^2(\ell+1)^2}\Big) \Bigg]\Bigg\} a^2\omega^2 +\Order{\alpha^2,a^3\omega^3}, \qquad H(L) \equiv \frac{2(L^2-m^2)(L^2-s^2)^2}{(2L-1)L^3(2L+1)}.
\nonumber
\end{eqnarray}
\end{widetext}
Here $\ell=\ell_i,\ell_i+1,\ell_i+2,\ldots$ is the multipole number, and we have $\ell_i\equiv\frac{1}{2}|m-s|+\frac{1}{2}|m+s|=\min(|m|,|s|)$.

The radial Teukolsky equation (\ref{eqn:Rs}),
\begin{eqnarray} \nonumber
&&\Bigg\{ \
\Delta_r^{-s}\frac{d}{dr}\Delta_r^{s+1}\frac{d}{dr}
+\frac{1}{\Delta_r}\left[ (1+\alpha)^2 K^2
- \imo s(1+\alpha)K \frac{d\Delta_r}
{dr} \right]
\\  && \qquad \label{eqn:Rr}
+4is(1+\alpha)\omega r -\frac{2\Lambda}{3}(s+1)(2s+1) r^2
\\  && \qquad\nonumber
+2s(1-\alpha) -\lambda \ \Bigg\} R = 0,
\end{eqnarray}
has five regular singularities: $r=r_c$, $r=r_e$, $r=r_i$, $r=r_0=-(r_c+r_e+r_i)$, and $r=\infty$.
In order to introduce the Frobenius expansion with the coefficients, satisfying the three-terms recurrence relation, we need to take into account behaviour of the solution in all the singular points.
The appropriate expansion is obtained in \cite{Yoshida:2010zzb}:
\begin{eqnarray}\nonumber
R_s(r)&=&\left(\frac{r-r_e}{r-r_0}\right)^{B_1}\left(\frac{r-r_i}{r-r_0}\right)^{B_2}\left(\frac{r_c-r}{r-r_0}\right)^{B_3}\times
\\\nonumber&&\times(r-r_0)^{-1-2s}\sum_{k=0}^{\infty}b_k\left(\frac{r-r_e}{r-r_0}\frac{r_c-r_0}{r_c-r_e}\right)^k.
\end{eqnarray}
The exponents $B_1$, $B_2$, $B_3$ and the corresponding recurrence relation coefficients can be found in the ancillary Mathematica\textregistered{} notebook \cite{acillary}.


\begin{thebibliography}{99}
\bibitem{Kokkotas:1999bd}
K.~D.~Kokkotas and B.~G.~Schmidt,
Living Rev. Rel. \textbf{2} (1999), 2
[arXiv:gr-qc/9909058 [gr-qc]].

\bibitem{Konoplya:2011qq}
R.~A.~Konoplya and A.~Zhidenko,
Rev. Mod. Phys. \textbf{83} (2011), 793-836
[arXiv:1102.4014 [gr-qc]].

\bibitem{Mashhoon}
B.~Mashhoon, in ``Proceedings of the Third Marcell Grossmann Meeting on General Relativity'', ((North-Holland Pub. Co., New York, 1983) p.599.

\bibitem{Ferrari:1984zz}
V.~Ferrari and B.~Mashhoon,
Phys. Rev. D \textbf{30}, 295-304 (1984).

\bibitem{Zhidenko:2003wq}
A.~Zhidenko,
Class. Quant. Grav. \textbf{21}, 273-280 (2004)
[arXiv:gr-qc/0307012 [gr-qc]].

\bibitem{Suzuki:1998vy}
H.~Suzuki, E.~Takasugi and H.~Umetsu,
Prog. Theor. Phys. \textbf{100}, 491-505 (1998)
[arXiv:gr-qc/9805064 [gr-qc]].

\bibitem{Konoplya:2007zx}
R.~A.~Konoplya and A.~Zhidenko,
Phys. Rev. D \textbf{76}, no.8, 084018 (2007)
[erratum: Phys. Rev. D \textbf{90}, no.2, 029901 (2014)]
[arXiv:0707.1890 [hep-th]].

\bibitem{Yoshida:2010zzb}
S.~Yoshida, N.~Uchikata and T.~Futamase,
Phys. Rev. D \textbf{81}, 044005 (2010).

\bibitem{Suzuki:1999nn}
H.~Suzuki, E.~Takasugi and H.~Umetsu,
Prog. Theor. Phys. \textbf{102}, 253-272 (1999)
[arXiv:gr-qc/9905040 [gr-qc]].

\bibitem{Panotopoulos:2020mii}
G.~Panotopoulos and Á.~Rincón,
Phys. Dark Univ. \textbf{31} (2021), 100743
[arXiv:2011.02860 [gr-qc]].

\bibitem{Churilova:2020mif}
M.~S.~Churilova,
Annals Phys. \textbf{427} (2021), 168425
[arXiv:2004.14172 [gr-qc]].

\bibitem{Tattersall:2018axd}
O.~J.~Tattersall,
Phys. Rev. D \textbf{98} (2018) no.10, 104013
[arXiv:1808.10758 [gr-qc]].

\bibitem{Dey:2018cws}
S.~Dey and S.~Chakrabarti,
Eur. Phys. J. C \textbf{79} (2019) no.6, 504
[arXiv:1807.09065 [gr-qc]].

\bibitem{Panotopoulos:2018hua}
G.~Panotopoulos,
Mod. Phys. Lett. A \textbf{33} (2018) no.23, 1850130
[arXiv:1807.03278 [gr-qc]].

\bibitem{Jansen:2017oag}
A.~Jansen,
Eur. Phys. J. Plus \textbf{132} (2017) no.12, 546
doi:10.1140/epjp/i2017-11825-9
[arXiv:1709.09178 [gr-qc]].

\bibitem{Breton:2017hwe}
N.~Bret\'on, T.~Clark and S.~Fernando,
Int. J. Mod. Phys. D \textbf{26} (2017) no.10, 1750112
[arXiv:1703.10070 [gr-qc]].

\bibitem{Cuyubamba:2016cug}
M.~A.~Cuyubamba, R.~A.~Konoplya and A.~Zhidenko,
Phys. Rev. D \textbf{93} (2016) no.10, 104053
[arXiv:1604.03604 [gr-qc]].

\bibitem{Fernando:2016ftj}
S.~Fernando,
Gen. Rel. Grav. \textbf{48} (2016) no.3, 24
[arXiv:1601.06407 [gr-qc]].

\bibitem{Zhang:2014xha}
Y.~Zhang, E.~K.~Li and J.~L.~Geng,
Gen. Rel. Grav. \textbf{46} (2014) no.10, 1728

\bibitem{Konoplya:2004uk}
R.~A.~Konoplya and A.~Zhidenko,
JHEP \textbf{06} (2004), 037
[arXiv:hep-th/0402080 [hep-th]].

\bibitem{Konoplya:2007jv}
R.~A.~Konoplya and A.~Zhidenko,
Nucl. Phys. B \textbf{777} (2007), 182-202
[arXiv:hep-th/0703231 [hep-th]].

\bibitem{Giammatteo:2005vu}
M.~Giammatteo and I.~G.~Moss,
Class. Quant. Grav. \textbf{22} (2005), 1803-1824
[arXiv:gr-qc/0502046 [gr-qc]].

\bibitem{Hod:2018lmi}
S.~Hod,
Phys. Lett. B \textbf{780} (2018), 221-226
[arXiv:1803.05443 [gr-qc]].

\bibitem{Cardoso:2017soq}
V.~Cardoso, J.~L.~Costa, K.~Destounis, P.~Hintz and A.~Jansen,
Phys. Rev. Lett. \textbf{120} (2018) no.3, 031103
[arXiv:1711.10502 [gr-qc]].

\bibitem{Dias:2018etb}
O.~J.~C.~Dias, H.~S.~Reall and J.~E.~Santos,
JHEP \textbf{10} (2018), 001
[arXiv:1808.02895 [gr-qc]].

\bibitem{Cardoso:2003sw}
V.~Cardoso and J.~P.~S.~Lemos,
Phys. Rev. D \textbf{67}, 084020 (2003)
[arXiv:gr-qc/0301078 [gr-qc]].

\bibitem{Molina:2003ff}
C.~Molina,
Phys. Rev. D \textbf{68} (2003), 064007
[arXiv:gr-qc/0304053 [gr-qc]].

\bibitem{Cardona:2017scd}
A.~F.~Cardona and C.~Molina,
Class. Quant. Grav. \textbf{34} (2017) no.24, 245002
[arXiv:1711.00479 [gr-qc]].

\bibitem{Jing:2003wq}
J.~l.~Jing,
Phys. Rev. D \textbf{69}, 084009 (2004)
[arXiv:gr-qc/0312079 [gr-qc]].

\bibitem{Chen:2005rm}
S.~B.~Chen and J.~L.~Jing,
Class. Quant. Grav. \textbf{22}, 1129-1141 (2005).

\bibitem{Chang:2005ki}
J.~F.~Chang and Y.~G.~Shen,
Nucl. Phys. B \textbf{712}, 347-370 (2005)
doi:10.1016/j.nuclphysb.2005.01.043
[arXiv:gr-qc/0502083 [gr-qc]].

\bibitem{Jing:2005dt}
J.~l.~Jing,
Phys. Rev. D \textbf{71}, 124006 (2005)
doi:10.1103/PhysRevD.71.124006
[arXiv:gr-qc/0502023 [gr-qc]].

\bibitem{Jing:2005pk}
J.~l.~Jing and Q.~y.~Pan,
Nucl. Phys. B \textbf{728}, 109-120 (2005)
doi:10.1016/j.nuclphysb.2005.08.038
[arXiv:gr-qc/0506098 [gr-qc]].

\bibitem{Jing:2005cb}
J.~l.~Jing,
JHEP \textbf{12}, 005 (2005)
doi:10.1088/1126-6708/2005/12/005
[arXiv:gr-qc/0512015 [gr-qc]].

\bibitem{Li:2013fka}
J.~Li, M.~Hong and K.~Lin,
Phys. Rev. D \textbf{88}, 064001 (2013)
[arXiv:1308.6499 [gr-qc]].

\bibitem{Varghese:2014xaa}
N.~Varghese and V.~C.~Kuriakose,
Mod. Phys. Lett. A \textbf{29}, 1450113 (2014)
[arXiv:1407.6292 [gr-qc]].

\bibitem{Wahlang:2017zvk}
W.~Wahlang, P.~A.~Jeena and S.~Chakrabarti,
Int. J. Mod. Phys. D \textbf{26}, no.14, 1750160 (2017)
[arXiv:1703.04286 [gr-qc]].

\bibitem{Blazquez-Salcedo:2017bld}
J.~L.~Bl\'azquez-Salcedo and C.~Knoll,
Phys. Rev. D \textbf{97}, no.4, 044020 (2018)
[arXiv:1709.07864 [gr-qc]].

\bibitem{Gonzalez:2018xrq}
P.~A.~Gonzalez, Y.~Vasquez and R.~N.~Villalobos,
Phys. Rev. D \textbf{98}, no.6, 064030 (2018)
[arXiv:1807.11827 [gr-qc]].

\bibitem{Zinhailo:2019rwd}
A.~F.~Zinhailo,
Eur. Phys. J. C \textbf{79}, no.11, 912 (2019)
[arXiv:1909.12664 [gr-qc]].

\bibitem{Wongjun:2019ydo}
P.~Wongjun, C.~H.~Chen and R.~Nakarachinda,
Phys. Rev. D \textbf{101}, no.12, 124033 (2020)
[arXiv:1910.05908 [gr-qc]].

\bibitem{Aragon:2020teq}
A.~Arag\'on, R.~B\'ecar, P.~A.~Gonz\'alez and Y.~V\'asquez,
Phys. Rev. D \textbf{103}, no.6, 064006 (2021)
[arXiv:2009.09436 [gr-qc]].

\bibitem{Chen:2021rty}
C.~H.~Chen, H.~T.~Cho, A.~Chrysostomou and A.~S.~Cornell,
Phys. Rev. D \textbf{104}, no.2, 024009 (2021)
[arXiv:2103.07777 [gr-qc]].

\bibitem{Shu:2005qv}
F.~W.~Shu and Y.~G.~Shen,
Phys. Lett. B \textbf{614}, 195-200 (2005)
[arXiv:gr-qc/0505161 [gr-qc]].

\bibitem{Zhang:2006bc}
Y.~Zhang and J.~L.~Jing,
Int. J. Mod. Phys. D \textbf{15}, 905-915 (2006).

\bibitem{Regge:1957td}
T.~Regge and J.~A.~Wheeler,
Phys. Rev. \textbf{108}, 1063-1069 (1957).

\bibitem{Brill:1957fx}
D.~R.~Brill and J.~A.~Wheeler,
Rev. Mod. Phys. \textbf{29}, 465-479 (1957).

\bibitem{Khanal:1983vb}
U.~Khanal,
Phys. Rev. D \textbf{28}, 1291-1297 (1983).


\bibitem{Newman:1961qr}
E.~Newman and R.~Penrose,
J. Math. Phys. \textbf{3}, 566-578 (1962).

\bibitem{Leaver:1985ax}
E.~W.~Leaver,
Proc. Roy. Soc. Lond. A \textbf{402}, 285-298 (1985).

\bibitem{Nollert:1993zz}
H.~P.~Nollert,
Phys. Rev. D \textbf{47}, 5253-5258 (1993).

\bibitem{Schutz:1985km}
B.~F.~Schutz and C.~M.~Will,
Astrophys. J. Lett. \textbf{291}, L33-L36 (1985).

\bibitem{Iyer:1986np}
S.~Iyer and C.~M.~Will,
Phys. Rev. D \textbf{35}, 3621 (1987).

\bibitem{Konoplya:2003ii}
R.~A.~Konoplya,
Phys. Rev. D \textbf{68}, 024018 (2003)
[arXiv:gr-qc/0303052 [gr-qc]].

\bibitem{Matyjasek:2017psv}
J.~Matyjasek and M.~Opala,
Phys. Rev. D \textbf{96}, no.2, 024011 (2017)
[arXiv:1704.00361 [gr-qc]].

\bibitem{Hatsuda:2019eoj}
Y.~Hatsuda,
Phys. Rev. D \textbf{101}, no.2, 024008 (2020)
[arXiv:1906.07232 [gr-qc]].

\bibitem{Konoplya:2019hlu}
R.~A.~Konoplya, A.~Zhidenko and A.~F.~Zinhailo,
Class. Quant. Grav. \textbf{36}, 155002 (2019)
[arXiv:1904.10333 [gr-qc]].

\bibitem{acillary}
The ancillary file with the three-terms recurrence relation coefficients for the Kerr-Newman-de Sitter black hole is available from \href{https://arxiv.org/src/2108.04858v1/anc/Kerr-Newman-dS-three-terms.nb}{https://arxiv.org/src/2108.04858v1/anc/}.

\bibitem{Gibbons:1993hg}
G.~W.~Gibbons and A.~R.~Steif,
Phys. Lett. B \textbf{314}, 13-20 (1993)
[arXiv:gr-qc/9305018 [gr-qc]].

\bibitem{Das:1996we}
S.~R.~Das, G.~W.~Gibbons and S.~D.~Mathur,
Phys. Rev. Lett. \textbf{78}, 417-419 (1997)
[arXiv:hep-th/9609052 [hep-th]].

\bibitem{Lopez-Ortega:2004vrh}
A.~Lopez-Ortega,
Gen. Rel. Grav. \textbf{36}, 1299-1319 (2004).

\bibitem{Chen:2019rob}
C.~H.~Chen, H.~T.~Cho, A.~S.~Cornell and G.~E.~Harmsen,
Phys. Rev. D \textbf{100}, no.10, 104018 (2019)
[arXiv:1907.11856 [gr-qc]].

\bibitem{Chen:2015jga}
C.~H.~Chen, H.~T.~Cho, A.~S.~Cornell, G.~Harmsen and W.~Naylor,
Chin. J. Phys. \textbf{53}, 110101 (2015)
[arXiv:1504.02579 [gr-qc]].

\bibitem{Teukolsky:1974yv}
S.~A.~Teukolsky and W.~H.~Press,
Astrophys. J. \textbf{193}, 443-461 (1974).


\end{thebibliography}
\end{document}